\documentstyle[11pt,epsfig]{article}

\newcommand{\beq}{\begin{equation}}
\newcommand{\eeq}{\end{equation}}
\newcommand{\bea}{\begin{eqnarray}}
\newcommand{\eea}{\end{eqnarray}}

\def\laq{~\raise 0.4ex\hbox{$<$}\kern -0.8em\lower
0.62 ex\hbox{$\sim$}~}
\def\gaq{~\raise 0.4ex\hbox{$>$}\kern -0.7em\lower
0.62 ex\hbox{$\sim$}~}

\def \pa {\partial}
\def \ti {\widetilde}
\def \ra {\rightarrow}
\def \la {\lambda}

\def \Da {\Delta}
\def \b {\beta}
\def \a {\alpha}

\def \sg {\sigma}
\def \da {\delta}

\def \r {\rho}

\def \Om {\Omega}

\begin{document}
\begin{titlepage}

\begin{flushright}
BA-TH/04-478\\
February 2004\\
gr-qc/0405083
\end{flushright}

\vspace*{1cm}

\begin{center}
\Huge{Towards a future singularity?}

\vspace*{1cm}

\large{M. Gasperini}

\bigskip
\normalsize

{\sl Dipartimento di Fisica,
Universit\`a di Bari, \\
Via G. Amendola 173, 70126 Bari, Italy\\
and\\
Istituto Nazionale di Fisica Nucleare, Sezione di Bari, Bari, Italy \\
\vspace{0.3cm}
E-mail: {\tt maurizio.gasperini@ba.infn.it}}

\vspace*{1.5cm}

\begin{abstract}
We discuss whether the future extrapolation of the present
cosmological state may lead to a singularity even in case of ``conventional" (negative) pressure of the dark energy field, namely $w=p/\rho \geq -1$. The discussion is based on an often neglected aspect of scalar-tensor models of gravity: the fact that different test particles may follow the geodesics of different metric frames, and the need for a frame-independent regularization of curvature singularities.
\end{abstract}

\end{center}

\vspace{1.5cm}
\begin{center}
---------------------------------------------\\
\vspace {5 mm}
{\sl Essay written for the {\bf 2004 
Awards for Essays on Gravitation}}\\
{\sl (Gravity Research Foundation, Wellesley Hills, MA, 02481-0004,
USA).}\\
\vspace{0.5 cm}
{\sl Selected for {\bf Honorable Mention}.}\\
 
\end{center}
\end{titlepage}

There is now increasing evidence that the present Universe is in a state of ``late-time" inflation, dominated by a cosmic ``dark energy" field with negative (enough) pressure \cite{1}. It is also well known that the accelerated evolution of the scale factor corresponds to a constant or decreasing Hubble scale ($\dot H \leq 0$) if the effective pressure of the gravitational sources satisfies $-1 \leq p/\r \leq -1/3$, and to a growing Hubble scale if $p/\r  <-1$. Both possibilities seem to be compatible with present observations, as a recent study of the parameter $w \equiv p/\r$ for the dominant dark energy component (including all data from CMBR, SNIa, large scale structure and Hubble space telescope) has restricted $w$ to be in the range $-1.38 < w < -0.82$, at the $95 \%$ conficence level \cite{2} (see also \cite{3,3a} for previous analyses with similar results). 

Models of dark energy with ``supernegative" equation of state ($w<-1$) may be based on {\em phantom} \cite{4} or {\em $k$-essence} \cite{5} fields, and are characterized by a growing curvature scale, thus describing a Universe evolving towards a future singularity \cite{3a}. It should be emphasized, however, that a future singularity is not excluded even in the more conventional (and, probably more realistic) case in which $w \geq -1$. This possibility should be taken into account, in particular, in the context of all ``quintessential" models of dark energy \cite{5a} in which the source of the cosmic acceleration is a scalar field, rolling down  (or up) an appropriate potential. 

A well known aspect of scalar-tensor models of gravity is indeed the freedom of introducing different ``frames" (i.e. different sets of fields parametrizing the same effective action), which are not kinematically  equivalent if the scalar field is not trivially constant. If, in addition, the scalar field is not universally coupled to all kinds of matter (this is typically what happens, for instance, when considering the string-theory dilaton \cite{6}), it is then possible that different test particles  follow the geodesics of different metric frames \cite{7}. 

The regularity/singularity properties of a geometric manifold, on the other hand, are fully determined by its geodesic structure (the absence of singularities, in particular, is associated to the property of ``geodesic completeness" \cite{8}). If the divergences of the curvature invariants are smoothed out in a given frame, but {\em not}   in all frames (examples of this kind are well known in a string cosmology context \cite{6,9}), then the singularities may in principle 
disappear only for a given class of test particles, while they persist, in practice, for other types of test particles. 

Such an event may typically occur in the context of cosmological solutions of scalar-tensor gravity characterized by a time-dependent scalar field $\phi$. The transformation of the curvature from one frame to another is indeed controlled by an appropriate function of $\phi$: it is then possible for the transformed curvature to blow up, at late times, even starting with a curvature which is always regular in the original frame. Most dark-energy models, on the other hand, are characterized by a rolling scalar component of gravity, thus motivating the question raised by the title of this paper. 

We will consider here a very simple scalar-tensor model of dark energy containing two different types of gravitational sources: i) ordinary (baryonic?) matter (and radiation), minimally coupled to gravity, and following the geodesics of the usual Einstein frame (E-frame) geometry;  ii) a more exhotic (dark matter?) field, dubbed 
X-field, non-minimally coupled to $\phi$, and following the geodesics of a different (X-frame) geometry. By using present observations to constrain the parameters of the E-frame geometry and of the dark energy field $\phi$, we will discuss the possible occurrence of future X-frames singularities, felt by X-matter. 

The model is based on the following E-frame action,
\bea
&&
S= \int d^4x \sqrt{-g} \left[-R +{1\over 2} (\nabla \phi)^2 -V(\phi)\right]
+ S_X+S_m, 
\nonumber\\&&
S_X= {1\over 2}  \int d^4x \sqrt{-g}~ (\nabla X)^2 e^{q_0\phi},
\label{1}
\eea
where we have used units $16 \pi G=1$, we have adopted a scalar-field representation of X-matter (coupled to $\phi$ with scalar charge $q_0$), and where the action $S_m$ describes ordinary matter (uncoupled to $\phi$ and minimally coupled to gravity). In this frame, the X-field satisfies the equation of motion
\beq
\nabla^2 X +q_0 \nabla_\mu \phi \nabla^\mu X=0,
\label{2}
\eeq
and the stress tensor $\sqrt{-g}~T_{\mu\nu}(X)=2 \da S_X/\da g^{\mu\nu}$ is not covariantly conserved since, from Eq. (2),
\beq
\nabla_\nu T_\mu^\nu(X)= -{1\over 2}q_0 \nabla_\mu \phi~
 (\nabla X)^2 e^{q_0\phi},  ~~~~~~~~ 
T_{\mu\nu}(X)= e^{q_0\phi} \left[ \nabla_\mu X  \nabla_\nu X -{1\over 2} g_{\mu\nu} ( \nabla X)^2 \right].
\label{3}
\eeq

More generally, given an X-field with scalar charge density 
$\sqrt{-g}\sg=-2 \da S_X/\da \phi$, the associated stress tensor satisfies the covariant equation \cite{10} $\nabla_\nu 
 T_\mu^\nu(X)=(\sg/2) \nabla_\mu \phi$. In any case, the lack of covariant conservation of the metric stress tensor ($ \nabla T \not=0$) implies that the world-line of a test X-particles corresponds to a non-geodesic path of the E-frame metric, even to lowest order (as shown by the multipole expansion of Eq. (\ref{3}) around the center of mass of a test particle made of X-matter \cite{10}). 

The motion of the X-field, however, is geodesic with respect to a different metric frame $\ti g$ such that
\beq
g_{\mu\nu}= \ti g_{\mu\nu} e^{-q_0 \phi},
\label{4}
\eeq
where the X-field is minimally and canonically coupled to the background geometry, 
\beq
S_X= {1\over 2}  \int d^4x \sqrt{-\ti g} ~(\ti \nabla X)^2. 
\label{5}
\eeq
In such a frame the metric stress tensor is covariantly conserved, thanks to the X-field equation of motion,
\beq
\ti \nabla^2 X=0, ~~~~~~ \ti \nabla_\nu \ti T_\mu^\nu(X)=0, ~~~~~~ 
\sqrt{-\ti g}~\ti T_{\mu\nu}(X)=2 \da S_X/\da \ti g^{\mu\nu},
\label{6}
\eeq
and the multipole expansion of the conservation equation 
$\ti  \nabla \ti T=0$ leads indeed, to leading order, to a geodesic path of the X-frame metric $\ti g$. 

For an isotropic and spatially flat cosmology, in particular, the curvature scale is controlled by the time evolution of the Hubble parameter and, using Eq. (\ref{4}) to transform scale factor and cosmic time between the two frames, one easily obtains: 
\bea
&&
\ti a = a e^{q_0\phi/2}, ~~~~~~~~~ d \ti t = dt e^{q_0\phi/2}, ~~~~~~~~~~~
H= {d \ln a \over dt}, ~~~~~~ \dot \phi = {d\phi \over dt}, \nonumber\\
&&
\ti H_X(t)= {d \ln \ti a \over d \ti t}=\left(H+ {q_0\over 2} \dot \phi\right) e^{-q_0\phi/2}.
\label{7}
\eea
In the last equation the X-frame Hubble scale is directly expressed in terms of the E-frame cosmic time $t$, and of the E-frame functions $H(t), \dot \phi(t), \phi(t)$. The behaviour of the X-frame curvature is thus immediately fixed by Eq. (\ref{7}) for any given configuration solving the E-frame cosmological equations. We can easily ask, in particular, whether $\ti H_X$ is growing even if the E-frame Hubble scale $H$ is decreasing. 

As a first, simple example we will discuss the case in which the X-field does not represent a dominant source of the cosmic (E-frame) expansion, in which ordinary matter is asymptotically negligible, and the Universe is dominated by the dark energy field with exponential potential, $V(\phi)= V_0 e^{-\la\phi}$. It is well known, with such a potential, that  a homogeneous scalar field behaves like a perfect fluid with equation of state $p/\r\equiv (\dot \phi^2/2-V)/(\dot \phi^2/2+V)=w=$ const, in a configuration described by the following particular exact solution of the E-frame equations:
\beq
a=t^\b, ~~~~ \phi= \phi_0 + \a \ln t, ~~~~
\b= {2\over 3(1+w)}, ~~~~ \a = {2\over \la} = \pm \left[8\over 3(1+w)\right]^{1/2},
\label{8}
\eeq
where the parameters $w$ and $\phi_0$ are related to $\la$ and $V_0$ by
\beq
w= {2\over 3} \la^2 -1, ~~~~~~~~~~~~ V_0 e^{-\la \phi_0}={4\over 3} { 1-w\over (1+w)^2}.
\label{9}
\eeq

This solution is valid for $\a,\la>0$ or $\a,\la<0$ (in both cases the potential energy $V(\phi)$ decreases in time like $t^{-2}$). With a suitable choice of $\la$, in particular, it is possible to obtain $w$ in the range $\{-1, -1/3\}$, corresponding to a phase of accelerated expansion and decreasing curvature ($ \dot a >0, \ddot a >0, \dot H <0$),  typical of the present Universe. 
For such an E-frame configuration we obtain, from Eq. (\ref{7}), 
$\ti H_X \sim t^{-q_0 \a /2 -1}$, so that $\ti H_X$ may be growing as $t\ra + \infty$ (even if $H$ is decreasing) provided $\a$ and $q_0$ have the opposite sign, and provided $1+q_0 \a/2 <0$, namely 
\beq
q_0^2 >3(1+w)/2.
\label {10}
\eeq

As an example of this possibility we should mention, for $\a >0$, the case in which the dark energy field is the dilaton \cite{11,12}, and the X-field is a zero-form of the Neveu-Scharwz (NS-NS) sector of superstring/supergravity theories \cite{13}, exponentially coupled to the dilaton through the (possibly loop-corrected) String-frame factor $e^{-k \phi}$.  The E-frame transformed coupling is $e^{(1-k)\phi}$, corresponding to the E-frame charge $q_0=1-k$. For $k>1$ we then obtain a negative charge $q_0$, possibly satisfying Eq. (\ref{10}) for values of $w$ close enough to $-1$ (as suggested by present observations). For the opposite case $\a<0$ we should mention the Kalb-Ramond axion associated to the NS-NS two form $B_{\mu\nu}$ \cite{6}: the String-frame coupling to the dilaton ($e^\phi$) corresponds to the E-frame coupling $e^{2\phi}$, and then to the charge $q_0=2$, which is positive and always satisfies the condition (\ref{10}). 

In both cases 
the X-frame curvature diverges in the limit in which the E-frame cosmic time $t$ goes to $+\infty$, but the X-matter field will reach the singularity of the X-frame metric in a {\em finite} proper-time interval, 
\beq
\Da \ti t = \int _{t_0}^\infty dt~ e^{q_0 \phi(t)/2} \sim 
 \int _{t_0}^\infty dt~ t^{q_0 \a/2} \sim t_0^{1+q_0\a/2} <\infty,
\label{11}
\eeq
as guaranteed by the condition $1+q_0\a/2<0$, required for the growth of $\ti H_X$. The situation is similar to that of the ``Big Rip" scenario \cite{3a}, with the difference that the gravitational explosion of X-matter will not affect the rest of the Universe because, in this example, the energy density of the X-field is not a dominant source of the cosmic background. 

As a second --and perhaps more interesting-- example we may thus consider the case in which the X-field is a dominant component of dark matter. In such a case the value of the scalar charge is no longer a free parameter, as it controls (toghether with the slope of the potential) the rate of the cosmic acceleration \cite{12,14}. We may thus expect, in principle, a stronger connection between the behaviour of $H$ and $\ti H_X$. 

In order to identify the X-field with a dark matter component we will
add a mass term to $S_X$,
\beq
S_X= {1\over 2}  \int d^4x \sqrt{-g}~ \left[(\nabla X)^2e^{q_0\phi}-
m^2  X^2e^{q_1\phi}\right]
\label{12}
\eeq
and, in the homogeneous limit $X=X(t)$, we will consider the X-field
configuration satisfying
\beq
\dot X^2 e^{q_0\phi}= m^2   X^2e^{q_1\phi},
\label{13}
\eeq
for which the space part of the stress tensor is vanishing, $T_i^j(X)=-
p_X \da_i^j=0$ (as appropriate to ``dust" dark matter). Using the
above condition, the coupled equations for the $X$ and $\phi$
fields can be written in terms of perfect fluid variables, and 
describe the coupling of dark energy to a dust-matter fluid carrying the scalar charge $q=q_1-q_0$. The E-frame cosmological equations, after imposing Eq. (\ref{13}),
are indeed reduced to the form
\bea
&&
6H^2=\r_\phi+\r_X, ~~~~~~
4 \dot H + 6 H^2 =-p_\phi,
\nonumber\\
&&
\dot \r_\phi +3 H (\r_\phi +p_\phi)+{\pa V\over \pa \phi}+{1\over
2}(q_1-q_0)\r_X\dot \phi=0,
\nonumber\\
&&
\dot \r_X +3 H \r_X -{1\over
2}(q_1-q_0)\r_X\dot \phi=0,
\label{14}
\eea
where
\bea
&&
\r_\phi ={1\over 2} \dot \phi^2 +V, ~~~~~~~~~~~~~
p_\phi ={1\over 2} \dot \phi^2 -V,
\nonumber \\
&&
\r_X= {1\over 2} \dot X^2 e^{q_0\phi}+ {1\over 2} m^2 X^2 e^{q_1\phi}
=  \dot X^2 e^{q_0\phi}= m^2 X^2 e^{q_1\phi}.
\label{15}
\eea

Using again an exponential potential, $V=V_0 e^{-\la \phi}$, such a
system of coupled equations can be satisfied by an asymptotic
configuration in which $\r_\phi, \r_X, H^2$ scale in time in the same
way \cite{12,14}, so that kinetic and potential energies of the $X$ and
$\phi$ fields keep constant in critical units. The dark energy parameter
$w=p_\phi/\r_\phi$ is also frozen at a constant value. One obtains, in
particular,
\bea
&&
H^2 \sim \r_\phi \sim\r_X \sim \dot \phi^2 \sim V\sim a^{-6 {\la\over
q+ 2\la}}, ~~~~~~~~
q= q_1-q_0,
\nonumber \\
&&
\Om_\phi = {\r_\phi\over 6H^2}={6 +q (q+2\la)\over (q+2\la)^2}, ~~~~~~~~
~~~~~~~~~~ w=-{q (q+2\la)\over 6 +q (q+2\la)},
\label{16}
\eea
from which 
\beq
a \sim t^{q+2\la\over 3 \la}, ~~~~~~~~~
\phi \sim {2\over \la} \ln t, ~~~~~~~~~
{\ddot a\over a H^2} = {q-\la \over q+2\la}.
\label {17}
\eeq
A phase of accelerated expansion, compatible with present bounds on
$w$ and $\Om_\phi$, and with a decreasing Hubble parameter (no
future singularity in the E-frame metric), corresponds in particular to
the case in which $q$ and $\la$ have the same sign, and $|q|
>|\la|$. In such a case $w\geq -1$ and $\dot H \leq 0$. 

In this example, the X-frame in which the X-field evolves
geodetically (like decoupled dust matter, $\dot {\ti \r}_X +3 \ti H_X \ti
\r_X=0$) is still determined by a conformal transformation of the type
(\ref{4}), in which $q_0$ is replaced however by $q=q_1-q_0$. From Eq. (\ref{7})
and (\ref{17}) we then obtain 
\beq
\ti H_X (t) \sim t^{-{q_1-q_0\over \la} -1}. 
\label{18}
\eeq
Thus, if sign$\{q_1-q_0\}=$ sign$\{\la\}$, namely if we choose to match present observations consistently with a decreasing E-frame Hubble scale, it turns out that the X-frame scale  $\ti H_X$ is also decreasing as $t \ra + \infty$.  
This means that, in this second example, the geodesic structure
of the two frames is too strongly correlated to allow a different
singularity behaviour. 

In conclusion, we have presented examples in which the absence of a
future singularity for the E-frame metric  may or may not correspond
to the presence of a future singularity in other metric frames. In
general, the notion of singularity is frame-dependent: different types
of matter, flowing along different geodesic networks, will live in
manifolds with different singularity structure. ``Bad" matter is doomed
to the hell of a future singularity, ``good" matter is doomed to the
heaven of an infinite time evolution. Let us hope that our world is made
of the right stuff, following the geodesics of the regular frame! 

\section*{Acknowledgements}I wish to thank Luca Amendola for useful information on recent observational results. 



\begin{thebibliography}{99}
\newcommand{\bb}{\bibitem}

\bibitem{1}S. Perlmutter et al., Nature {\bf 391}, 51 (1998); A. G. Riess et al., Astron. J. {\bf 116}, 1009 (1998); 
P. de Bernardis et al.,  Nature {\bf 404}, 955 (2000); 
S. Hanay et al., Astrophys. J. Lett. {\bf 545}, L5 (2000); N. W. Alverson et al., Astrophys. J. {\bf 568}, 38 (2002);
D. N. Spergel et al., Astrophys. J. Suppl. {\bf 148}, 175 (2003). 

\bb{2}A. Melchiorri et al., Phys. Rev. {\bf D68}, 043509 (2003).

\bb{3}S. Hamestad and E. Morstel, Phys. Rev. {\bf D66}, 063508  (2002).

\bb{3a}R. R. Caldwell, M. Kamionkowski and N. N. Weinberg, Phys. Rev. Lett. {\bf 91}, 07130 (2003). 

\bb{4}R. R. Caldwell, Phys. Lett. {\bf B545}, 23 (2002).

\bb{5} C. Armendariz-Picon, V. Mukhanov and P. J. Steinhardt, 
 Phys. Rev. Lett. {\bf 85}, 4438 (2000). 

\bb{5a}B. Ratra and P. J. E. Peebles, Phys. Rev. {\bf D37}, 3406 (1988); C. Wetterich, Nucl. Phys. {\bf B302}, 668 (1988); M. S. Turner and C. White, Phys. Rev. {\bf D56}, 4439 (1997); 
R. R. Caldwell, R. Dave and P. J. Steinhardt,  Phys. Rev. Lett. {\bf 80}, 1582 (1998).

\bb{6} See for instance M. Gasperini and G. Veneziano, Phys. Rep. 
 {\bf 373}, 1 (2003).

\bb{7}See for instance N. Kaloper and K. A. Olive,  Phys. Rev. {\bf D57}, 811 (1998).

\bb{8}See for instance S. W.  Hawking  and G. F. R. Ellis, {\em The large scale structure of spacetime} (University Press, Cambridge, 1973). 

\bb{9}M. Gasperini and M. Giovannini, Phys. Lett. {\bf 287}, 56 (1992); M. Gasperini, M. Maggiore,
 and G. Veneziano, Nucl. Phys. {\bf B494}, 315  (1997).

\bb{10}M. Gasperini, Phys. Lett. {\bf 470}, 67 (1999). 

\bb{11} M. Gasperini, Phys. Rev.  {\bf D64}, 043510  
(2001). 

\bb{12}M. Gasperini, F. Piazza and G. Veneziano,   Phys. Rev. {\bf D65}, 023508 (2001).

\bb{13} See for instance J. E. Lidsey, D. Wands and E. J. Copeland,  Phys. Rep. {\bf 337}, 343 (2000).


\bb{14}L. Amendola,  Phys. Rev. {\bf D 62}, 043511(2000); 
L. Amendola and D. Tocchini-Valentini  Phys. Rev. {\bf D 64}, 043509 (2001); L. Amendola, M. Gasperini, D. Tocchini-Valentini 
and C. Ungarelli,  Phys. Rev. {\bf D 67}, 043512 (2003);
M. Gasperini, hep-th/0310293. 







\end{thebibliography}
\end{document}